\documentclass[runningheads]{llncs}
\usepackage{algorithm}
\usepackage{algpseudocode}

\usepackage{microtype}
\usepackage{graphicx}
\usepackage{subfigure}
\usepackage{booktabs} 
\usepackage{hyperref}

\usepackage{url}            
\usepackage{amsfonts}       
\usepackage{nicefrac}       
\usepackage{amsmath}
\usepackage{comment}
\usepackage{float}
\usepackage{enumitem}
\usepackage{array}
\usepackage{wrapfig}

\usepackage{color}
\usepackage{caption}

\usepackage{natbib}

\newcommand{\E}{\mathbb{E}}
\newcommand{\R}{\mathbb{R}}
\newcommand{\N}{\mathcal{N}}

\renewcommand{\cite}{\citep}

\newcommand{\Rs}{\mathcal{R}}

\begin{document}
\title{Sibling Regression for Generalized Linear Models}
%
%
\author{Shiv Shankar\inst{1} \and
Daniel Sheldon\inst{1,2} }
\authorrunning{Shivshankar and D. Sheldon}
%
\institute{ University of Massachusetts, Amherst, MA 01003, USA  \email{\{sshankar,sheldon\}@cs.umass.edu} \and
Mount Holyoke College, South Hadley, MA 01075, USA }
\maketitle              

\begin{abstract}
 Field observations form the basis of many scientific studies, especially in ecological and social sciences. Despite efforts to conduct such surveys in a standardized way, observations can be prone to systematic measurement errors. The removal of systematic variability introduced by the observation process, if possible, can greatly increase the value of this data.
Existing non-parametric techniques for correcting such errors assume linear additive noise models. This leads to biased estimates when applied to generalized linear models (GLM). We present an approach based on residual functions to address this limitation. We then demonstrate its effectiveness on synthetic data and show it reduces systematic detection variability in moth surveys.  

\end{abstract}

\section{Introduction}

\begin{figure}[htb]
\centering
\subfigure[]{
\includegraphics[width=1.05\columnwidth]{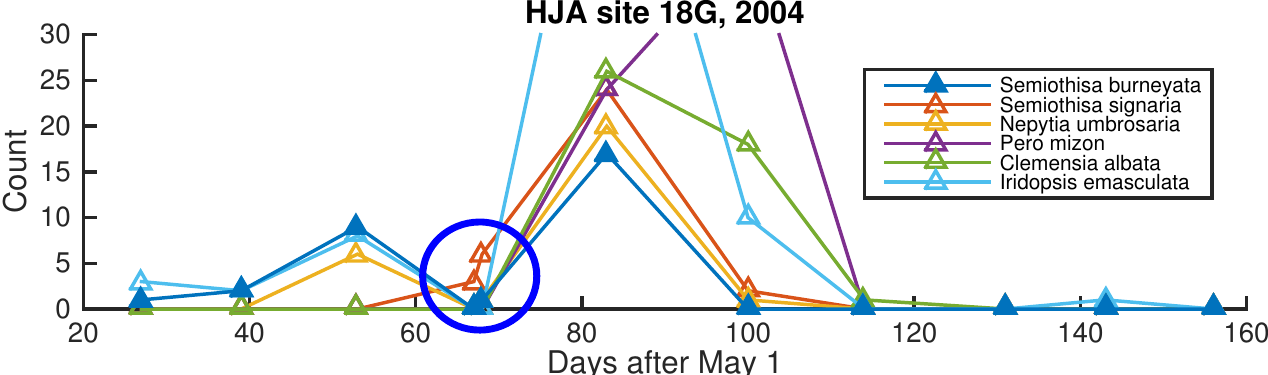}
}
\subfigure[]{
\includegraphics[width=1.05\columnwidth]{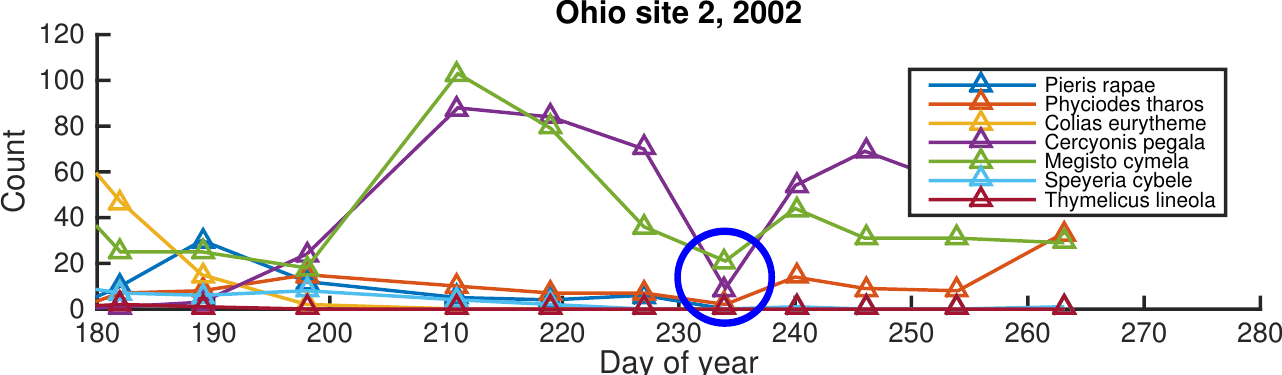}
}
\caption{Systematic detection error in moth and butterfly counts: (a) Correlated counts of other moths
  strongly suggest this is a detection problem. (b) Correlated
  detection errors in butterfly counts on day 234. }
\label{fig:detection}
\end{figure}

Observational data is increasingly important across a range of domains and may be affected by measurement error. Failure to account for systemic measurement error can lead to incorrect inferences.
Consider a field study of moth counts for estimating the abundance of different moth species over time. Figure~\ref{fig:detection}(a) shows the counts of \textit{Semiothisa burneyata} together with 5 other
species. We see that \textit{all} species had abnormally low counts on the same day. This suggests that the low count is due to a confounder and not an actual drop in the population. The same phenomenon is also prevalent in butterfly counts (Figure~\ref{fig:detection}(b)), where poor weather can limit detectability.

An abstract version of the aforementioned situation can be represented by Figure \ref{fig:sglm-sibling}. $X$ here represents ecological factors such as temperature, season etc; and $Z_1,Z_2$ represent the true abundance of species ( such as moths). $N$ is a corrupting noise (such as lunar phase) which affects the observable abundance $\theta$ of the organisms. $Y$ represents an observation of the population and is modeled as a sample drawn distribution of observed abundance ( for eg a Poisson distribution). Directly trying to fit a model ignoring the noise $N$ can lead to erroneous conclusions about key factors such as effect of temperature on the population. 

Distinguishing observational noise and measurement variability from true variability, often requires repeated measurements which in many cases can be expensive, if not impossible. 
However, sometimes even \textbf{\textit{ in the absence of repeated measurements}}, the effect of confounding noise can be estimated. Sibling regressions~\citep{Scholkopf15} refer to one such technique that use auxiliary variables influenced by a shared noise factor to estimate the true value of the variable of interest. These techniques work without any parametric assumption about the noise distribution ~\citep{Scholkopf15,3qs}. However, these works assume an additive linear noise model, which limits their applications. For example in the aforementioned insect population case the effect of noise is more naturally modeled as multiplicative rather than additive. 

We introduce a method that extends these ideas to generalized linear models (GLMs) and general exponential family models. First, we model non-linear effect of noise by considering it as an additive variable in the natural parameters of the underlying exponential family. Secondly, instead of joint inference of the latent variables, a stagewise approach is used. This stagewise approach is justified from a generalized interpretation of sibling regression. We provide justification behind our approach and then test it on synthetic data to quantitatively demonstrate the effectiveness of our proposed technique. 
Finally, we apply it to the moth survey data set used in~\citet{3qs} and show that it reduces measurement error more effectively than prior techniques.

\section{Related Work}
Estimation of observer variation can be framed as a causal effect estimation problem ~\citep{Bang05,Athey16}. Such conditional estimates often require that all potential causes of confounding errors have been measured~\citep{sharma18}. However, real-life observational studies are often incomplete, and hence these assumptions are unlikely to hold. 

~\citet{natarajandrt13,menon15} develop techniques to handle measurement error as a latent variable. Similar approaches have been used to model observer noise as class-conditional noise ~\citep{hutchinson2017species,yu2014latent}. One concern with such approaches is that they are generally unidentifiable. 

A related set of literature is on estimation with missing covariates \citep{jones1996indicator,little1992regression}. These are generally estimated via Monte-Carlo methods \citep{Ibrahim1992incomplete}, Expectation-maximization like methods \citep{ibrahimlc_99} or by latent class analysis \citep{formann1996latent}. Another set of approaches require strong parametric assumptions about the joint distribution \citep{little1992regression}. Like other latent variable models, these can be unidentifiable and often have multiple solutions \citep{horton1999maximum}.

\citet{MacKenzie2002} and \citet{Royle2004} learn an explicit model of the detection process to isolate observational error in ecological surveys using repeated measurements. Various identifiability criteria have also been proposed for such models ~\citep{solymos2016revisiting, knape2016assumptions}. ~\citet{lele2012dealing} extend these techniques to the case with only single observation. However, these models are only as reliable as the assumptions made about the noise variable. Our approach on the other hand makes weaker assumptions about the form of noise.

\citet{Scholkopf15} introduced 'Half-sibling regression'; an approach for denoising of independent variables. This approach is both identifiable and does not make assumptions about the prior distribution of noise. \citet{3qs} further extended the technique to the case when the variables of interest are only conditionally independent given an observed common cause.

\section{Preliminaries}
\label{sec:preliminaries}
\paragraph{Exponential Family Distributions}
A random variable $Y$ is said to be from an exponential family~\citep{Kupperman58} if its density can be written as

\begin{align*}
    p(y | \theta) = h(y)\exp(\theta^TT(y)-A(\theta))
\end{align*}
where $\theta$ are the (natural) parameters of the distribution, $T(y)$ is a function of the value $y$ called the \emph{sufficient statistic}, and $A(\theta)$ is the log normalizing constant, and $h(y)$ is a base measure. 

Given an exponential family density $p(y | \theta)$, let $L(y, \theta) = -\log p(y | \theta)$ be the negative log-likelihood loss function at data point $y$, and let $L(\theta) = \frac{1}{m}\sum_{i=1}^m L( y^{(j)}, \theta)$ be the overall loss of a sample $y^{(1)}, \ldots, y^{(m)}$ from the model. Also let $I(\theta) = \nabla^2 A(\theta)$ be the Fisher information. 

We summarize few standard properties ~\citep[e.g., see][]{kollerbook} of exponential families that we will be of use later:
\begin{enumerate}
\item $\nabla A(\theta) = \E[T(Y)]$.
\item $\nabla_\theta L(y, \theta) = A(\theta) - T(y)$
\item $\forall y:  \nabla^2_\theta L(y, \theta) = \nabla^2 A(\theta) = I(\theta)$. This also implies that $\nabla^2 L(\theta) = \nabla^2 A(\theta) = I(\theta)$.
\end{enumerate}

\paragraph{Generalized Linear Models}
Generalized Linear Models (GLMs)~\citep{Nelder72} are a generalization of linear regression models where the output variable is not necessarily Gaussian.
In a GLM, the conditional distribution of $Y$ given covariates $X$ is an exponential family with mean $\E[Y | X]$ is related to a linear combination of the covariates by the link function $g$, and with the identity function as the sufficient statistic, i.e., $T(Y)=Y$. We focus on the special case of GLMs with \emph{canonical link functions}, for which $\theta = X^T \beta$, i.e., the natural parameter itself is a linear function of covariates. For the canonical GLM, the following properties hold \citep{kollerbook} \footnote{These properties can be obtained by applying the aforementioned exponential family properties.
}:
\begin{enumerate}[topsep=2pt, itemsep=2pt, leftmargin=*]
\item The link function is determined by $A$ : $\E[Y|X] = g^{-1}(X^T\beta) = \nabla A(X^T\beta)$
\item The conditional Fisher information matrix $I_{Y|X} = I_{\cdot|X}(\theta) =  \nabla^2 A(X^T\beta)$
\end{enumerate}

GLMs are commonly used in many applications. For example, logistic regression for binary classification is identical to a Bernoulli GLM. Similarly, regressions where the response variable is of a count nature, such as the number of individuals of a species, are based on Poisson GLM.

\paragraph{Sibling Regression}
Sibling regression~\citep{Scholkopf15,3qs} is a technique which detects and corrects for confounding by latent noise variable using observations of another variable influenced by the same noise variable. 

\begin{figure*}[htb]
  \centering
    \subfigure[]{\label{fig:half-sibling}      \includegraphics[width=0.24\linewidth]{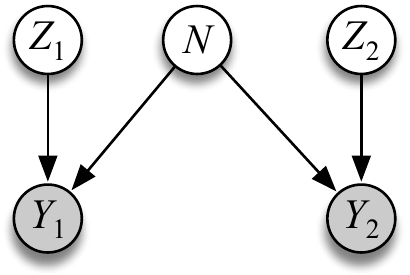}
    }
    \subfigure[]{\label{fig:three-quarter-sibling}
      \includegraphics[width=0.24\linewidth]{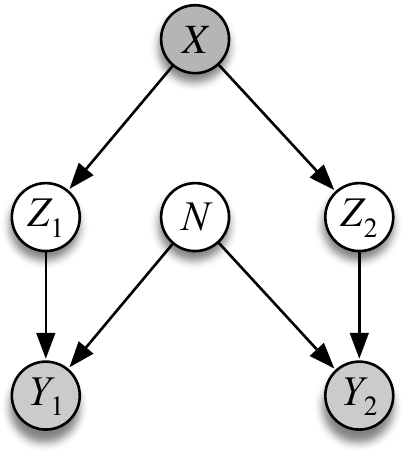}
    }
    \subfigure[]{\label{fig:glm-sibling} 
      \vspace{10pt}
      \includegraphics[width=0.24\linewidth]{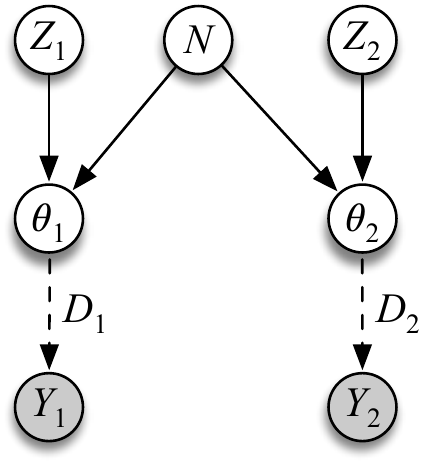}
    }
    \captionof{figure}{ \label{fig:sibling} (a) Half-sibling regression. (b) Three-quarter sibling regression. (c) Generalized linear sibling model without covariates.}
\end{figure*}

The causal models depicted in Figure~\ref{fig:sibling} capture the essential use case of sibling regression techniques.
Here, $Z_1, Z_2$ represent the unobserved variables of interest, which we would like to estimate using the observed variables $Y_1, Y_2$; however the observations are confounded by the common unobserved noise $N$. \citet{Scholkopf15} use the model illustrated in Figure ~\ref{fig:half-sibling} as the basis of their \emph{half-sibling regression} approach to denoise measurements of stellar brightness. 
The \textbf{\textit{a priori}} independence of $Z_1,Z_2$ implies that any correlation between $Y_1,Y_2$ is an artifact of the noise process.
%

The half-sibling estimator of $Z_1$ from~\citep{Scholkopf15} is
\begin{equation}
\hat{Z_1} = Y_1 - \E[Y_1|Y_2] + \E[Y_1]
\label{eqn:hls}
\end{equation}
This equation can be interpreted as removing the part of $Y_1$ that can be explained by $Y_2$, and then adding back the mean $\E[Y_1]$. \citet{Scholkopf15} did not include the final $\E[Y_1]$ term, so our estimator differs from the original by a constant; we include this term so that $\E[\hat{Z}_1] = \E[Y_1]$. In practice, the expectations required are estimated using a regression model to predict $Y_1$ from $Y_2$, which gives rise to the name ``sibling regression''.

~\textit{Three-quarter sibling} regression~\citep{3qs} generalizes the idea of half-sibling regression to the case when $Z_1$ and $Z_2$ are not independent but are conditionally independent given an observed covariate $X$. In the application considered by ~\citet{3qs}, these variables are population counts of different species in a given survey; the assumed dependence structure is shown in Figure~\ref{fig:three-quarter-sibling}. This estimator has a similar form to the half sibling version, except the expectations now condition on $X$:
\begin{equation}
  \hat{Z}_{1|X} = Y_1 - \E[Y_1|X, Y_2] + \E[Y_1 | X]
  \label{eqn:3qs}
\end{equation}


\section{Sibling Regression for Generalized Linear Models}

\paragraph{Model}
We now formally specify the model of interest.
We will focus first on the half-sibling style model with no common cause $X$, and revisit the more general case later.
Figure~\ref{fig:glm-sibling} shows the assumed independence structure. As in the previous models, the true variables of interest are $Z_1$ and $Z_2$, and the observed variables are $Y_1$ and $Y_2$. The variable $N$ is confounding noise that influences observations of both variables. Unlike half-sibling regression, exponential-family sampling distributions $D_1$ and $D_2$ mediate the relationship between the hidden and observed variables; the variables $\theta_1$ and $\theta_2$ are the parameters of these distributions. The dotted arrow indicate that the relation between $Y_i$ and $\theta_i$ is via sampling from an exponential family distribution, and not a direct functional dependency. On the other hand $\theta_1,\theta_2$ are deterministic functions of $(Z_1,N)$ and $(Z_2,N)$, respectively.
We assume the noise acts additively on the natural parameter of the exponential family. 
Mathematically the model is
\begin{equation}
Y_i \sim D_i(\theta_i), \quad \theta_i = Z_i + N \quad i \in \{1, 2\}.
\label{eqn:model}
\end{equation}
More generally, the noise term $N$ can be replaced by a non-linear function $\psi_i(N)$ mediating the relationship between the noise mechanism and the resultant additive noise; however, in general $\psi_i$ will not be estimable so we prefer to directly model the noise as additive.

The key idea behind sibling regression techniques is to find a ``signature'' of the latent noise variable using observations of another variable. In this section we will motivate our approach by reformulating prior sibling regression methods in terms of \emph{residuals}.

\begin{lemma}
 \label{clm:claim1}
  Let $R_i = Y_i - \E[Y_i]$ be the \emph{residual} (deviation from the mean) of $Y_i$ in the half-sibling model, and let $R_{i|X} = Y_i - \E[Y_i | X]$ be the residual relative to the conditional mean in the three-quarter sibling model. The estimators of Eqs.~\eqref{eqn:hls} and~\eqref{eqn:3qs} can be rewritten as
  $$
  \hat{Z}_1 = Y_1 - \E[R_1 | R_2], \quad \hat{Z}_{1|X} = Y_1 - \E\big[R_{1|X} \mid R_{2|X}\big].
  $$
\end{lemma}
\begin{proof}For the half-sibling case, write
\begin{align*}
    \hat{Z_1} &= Y_1 - \E[Y_1 \mid Y_2] + \E[Y_1] \\
    &= Y_1 - \E\big[Y_1 - \E[Y_1] \mid Y_2\big] \\
    &= Y_1 - \E\big[Y_1 - \E[Y_1] \mid Y_2 - \E[Y_2]\big] \\
    &= Y_1 - \E[R_1 \mid R_2] 
\end{align*}
The three-quarter case is similar. 
\end{proof}
This formulation provides a concise interpretation of sibling regressions as regressing the residuals of $Y_1$ on those of $Y_2$.

\subsection{Extension to GLM}

\noindent\begin{minipage}{\linewidth}
    \centering
     \begin{minipage}{0.4\linewidth}
    \includegraphics[width=0.85\linewidth]{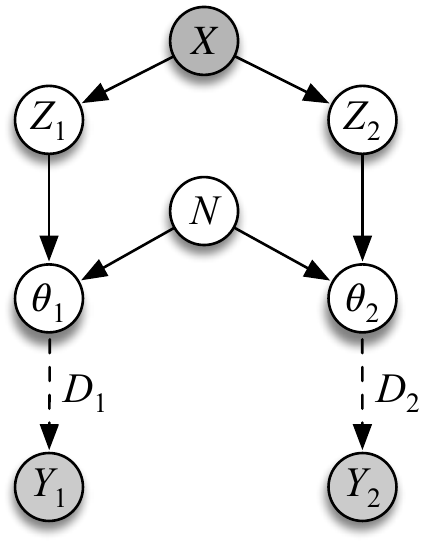}
  \captionof{figure}{Generalized sibling model with covariates \label{fig:sglm-sibling}}
      \end{minipage}
      \begin{minipage}{0.58\linewidth}
      \paragraph{Problem Statement}
      The problem is to obtain estimates of $Z_1$ and $Z_2$ given $m$ independent observations $(y_1^{(1)}, y_2^{(1)}), \ldots (y_1^{(m)}, y_2^{(m)})$ of $Y_1$ and $Y_2$. The general idea is, as in prior sibling regressions, to model and remove the signature of the noise variable. ~\citet{Scholkopf15,3qs} solve this problem in the special case where $D_i$ is Dirac. We wish to extend this to the case of other sampling distributions. 
      \end{minipage}
\end{minipage}

By symmetry, it suffices to focus on estimating only one variable, so we will henceforth consider $Z_1$ to be the estimation target.
We allow the possibility that $Z_2$ is multivariate to model the case when there are many siblings that each contain a trace of the noise variable.


Inspiring from the residual form of sibling-regression \ref{clm:claim1}, we derive a residual which can be used to approximate the confounding noise in the GLM case.

Computing a residual requires defining a reference, in analogy to the conditional global mean $\E[Y_1|X]$ used in Equation \ref{eqn:3qs}.
We will use the maximum-likelihood estimate under the ``global'' model $Y_1 \sim D_1(X^T\beta)$ as the reference. Let $\hat{\beta}$ be the estimated
regression coefficients for the relationship between the covariates $X$ and variable $Y_1$.
The corresponding model predictions $\hat{Y}_1$ are then given by $g^{-1}(X^T\hat{\beta})$ where $g$ is the link function. We define $\Rs$ as:

\begin{equation}
\Rs(Y) = I_{Y|X}^{-1}(Y - \hat{Y})
\label{eqn:sglm}
\end{equation}
where $I_{Y|X}$ is the conditional Fisher information.

\begin{proposition}
\begin{align*}
\E [\Rs(Y_1)| X, N] &= Z_1 + N - X^T\hat{\beta} + O(|Z_1 + N - X^T\hat{\beta}|^2) \\
& \approx Z_1 + N - X^T\hat{\beta}    
\end{align*}

\label{claim:resid}
\end{proposition}
\begin{proof}
We temporarily drop the subscript so that $(Y | Z, N) \sim D(\theta)$ where $\theta = Z + N$ is the natural parameter of the exponential family distribution $D$ corresponding to the specified GLM. 
$$
\begin{aligned}
\E [\Rs(Y)| X, N] 
=& \E\big[ I_{Y|X}^{-1}( Y - \hat{Y})| X, N]\\
=& \E\big[ [I_{Y|X}(\hat{\beta})]^{-1}( Y -  \nabla A(X^T\hat{\beta})| X, N]\\
=&  [I_{Y|X}(\hat{\beta})]^{-1} ( \E\big[Y| X, N] - \nabla A(X^T\hat{\beta})) \\
=&  [I_{Y|X}(\hat{\beta})]^{-1} ( \nabla A(Z + N) - \nabla A(X^T\hat{\beta})) \\
=&  [I_{Y|X}(\hat{\beta})]^{-1} [ \nabla^2 A(X^T\hat{\beta} )  (Z + N -X^T\hat{\beta}) \\
& \qquad +  O(|Z + N  -X^T\hat{\beta} |^2) ] \\
=&  [ (Z + N  - X^T\hat{\beta}) + O(|Z + N  - X^T\hat{\beta} |^2) ] \\
\end{aligned}
$$
The second line used the definition of canonical link function.
The third uses linearity of expectation.
The fourth line uses the first property of exponential families from Section~\ref{sec:preliminaries}.
The fifth line applies the Taylor theorem to $\nabla A$ about $X^T\hat{\beta}$. 
The last two lines use our previous definitions that $I_{Y|X} = \nabla^2 A(\theta)$ and $\theta =  X^T\beta + N $.
Restoring the subscripts on all variables except $N$, which is shared, gives the result.
\end{proof}

In simpler terms, $\Rs(Y_i)$ can provide a useful approximation for $Z_i + N - X^T\hat{\beta}_i$. Moreover, if we assume that $X$ largely explains $Z_i$, then the above expression is dominated by $N$. We would then like to use $\Rs$ to estimate and correct for the noise. 
\begin{proposition}
  Let the approximation in Proposition \ref{claim:resid} be exact, i.e.,
  $\E [\Rs(Y_i)| X, N] = Z_i + N - X^T\hat{\beta}_i$, then
  $\E[\Rs(Y_1)| \Rs(Y_2), X] - \E[\Rs(Y_1)| X]  = \E[N| X, \Rs(Y_2)] $ upto a constant.
\label{claim:reg}
\end{proposition}
The derivation is analogous to the one presented by \citet{3qs} and is given in Appendix \ref{apx:proof_reg}. While the higher order residual terms in Claim \ref{claim:resid} generally cannot be ignored, Claim \ref{claim:reg} informally justifies how we can estimate $N$ by regressing  $\Rs(Y_1)$ on $\Rs(Y_2)$.

Note that $\Rs(Y_i)$ is a \emph{random variable} and that the above expression is for the expectation of $\Rs$ which may not be exactly known (due to limited number of samples drawn from each conditional distribution) . However we observe that a) $Z_1,Z_2$ are independent conditional on the observed $X$ and b) the noise $N$ is independent of $X$ and $Z$ and is common between $\Rs(Y_1),\Rs(Y_2)$. This suggests that in presence of multiple auxiliary variables ( let $Y_{-i}$ denote all $Y$ variables except $Y_i$ ) , we can improve our estimate of $N$ by combining information from all of them. Since $Z_i$'s are conditionally independent (given $X$) of each other, we can expect their variation to cancel each other on averaging. The noise variable on the other hand being common will not cancel. Appendix \ref{apx:heuristic} provides a justification for this statement. %

Once we have an estimate $\hat{N}$ of the noise $N$, we can now estimate $Z_1$. However unlike standard sibling regression (Figures ~\ref{fig:half-sibling},~\ref{fig:three-quarter-sibling}), where the observed $Y_1$ was direct measurement of $Z_1$, in our case the relationship gets mediated via $\theta_1$. If we had the true values of $\theta_1$ one can directly use Equation \ref{eqn:3qs} to obtain $Z_1$. One can try to obtain $\theta_1$ from $Y_1$; but since $Y_1$ includes sampling error affecting our estimates of $Z_1$. Instead we rely once again on the fact that the model is a GLM, which are efficient and unbiased when all covariates are available. We re-estimate our original GLM fit but now with additional covariate $\hat{N}$ which is a proxy for $N$.

\paragraph{Implementation}
Based on the above insights, we propose Algorithm \ref{alg:calibrate}, labelled Sibling GLM (or SGLM) for obtaining denoised estimates from GLM models. In practice, the conditional expectations required are obtained by fitting regressors for the target quantity from the conditioning variables . As such we denote them by $\hat{E}[\cdot | \cdot]$):\; to distinguish them from true expectation values. Since, per Claim \ref{claim:resid}, the true expected value is approximately linear in the conditioning variables, in our experiments we used ordinary least squares regression to estimate the conditional expectations in Step 3.

\begin{algorithm}[t]
  \small{
  \textbf{Input:} $( X^{(k)}, Y_1^{(k)},
    Y_2^{(k)})_{k=1,\ldots, n}$ \\
  \textbf{Output:} Estimates of latent variable $\hat{Z}_1$\\
  \textbf{Training:} denote fitted regression models by $\hat{E}[\cdot | \cdot]$):\;
  \begin{enumerate}
  \item Compute $\hat{\E}[Y_1 | X], \hat{\E}[Y_2 | X]$ by training suitable GLM
  \item Compute $\Rs(Y_1),\Rs(Y_2)$ as given by Equation \ref{eqn:sglm}
  \item Fit regression models for $\Rs(Y_1)$ using $\Rs(Y_2), X$ as predictors to obtain estimators of $\hat{\E}[\Rs(Y_1) |  \Rs(Y_2), X], \hat{\E}[\Rs(Y_1) |  X]$
  \item Create $\hat{N}$ such that its $k^{\text{th}}$ value $ \hat{N}^{(k)} = \hat{\E}[\Rs(Y_1) |  \Rs(Y_2)^{(k)}, X^{(k)}] -
    \hat{\E}[\Rs(Y_1) | X^{(k)}]$ $ \text{ }  \forall k \in [1,n]$
  \item Estimate $\hat{Z}_1$ by fitting GLM models with $n$ as an additional covariate i.e $\hat{E}[Y_1 | X, n]$ 
  \end{enumerate}
    \caption{SGLM Algorithm}
    \label{alg:calibrate}
  }
\end{algorithm}


%
%

\section{Experiments}
In this section, we experimentally demonstrate the ability of
our approach to reduce estimation errors in our proposed setting, first using simulations and semi-synthetic examples and then with a moth survey data set.
Furthermore in our experiments we found the correlation between $\Rs(Y_i)$ and $X$ to be small, and therefore simplified Step 3 and 4 in Algorithm \ref{alg:calibrate} simplify to $N = \hat{\E}[\Rs(Y_1)|\Rs(Y_2)]$

\subsection{Synthetic Experiment}
We first conduct simulation experiments where, by design, the true value of $Z_1$ is known. We can then quantitatively measure the performance of our method across different ranges of available auxiliary information contained within $Y_{-1}$.

\textbf{Description}
We borrow the approach of \citet{Scholkopf15}, generalized to standard GLM distributions. We run these experiments for Poisson and Gamma distributions. This simulation was conducted by generating 120 observations of 20 different Poisson variables. Each individual observation was obtained via a noise-corrupted Poisson or Gamma distribution where, for each observation, the noise affected all variables simultaneously as described below. Specifically, each variable $Y_i$ at a single observation time is obtained via a generative process dependent on  $X \in \R$ and
noise variable $N \in \R$ as:
\[
  Y_i \sim D (\underbrace{w^{(i)}_X X}_{Z_i} + w^{(i)}_NN + \epsilon)
\]
The variables $X$ and $N$ are
drawn uniformly from $[-1, 1]$. Similarly, the coefficient $w^{(i)}_N$ is drawn from a uniform distribution on $[-1,1]$, while $\epsilon \sim \N(0, \sigma^2_\epsilon)$ is independent noise. Finally, $w^{(i)}_X$ is drawn from the standard conjugate prior for the distribution $D$.

\textbf{Results}
We conducted these simulations and measured the error in the estimated $Z$ versus the true $Z$. Due to inherent variability caused by sampling, the error will not go down to zero. We present below the results on Poisson regression, while other results can be found in the appendix.


\begin{figure*}[t]
\centering

\subfigure[]{
  \includegraphics[width=0.4\textwidth]{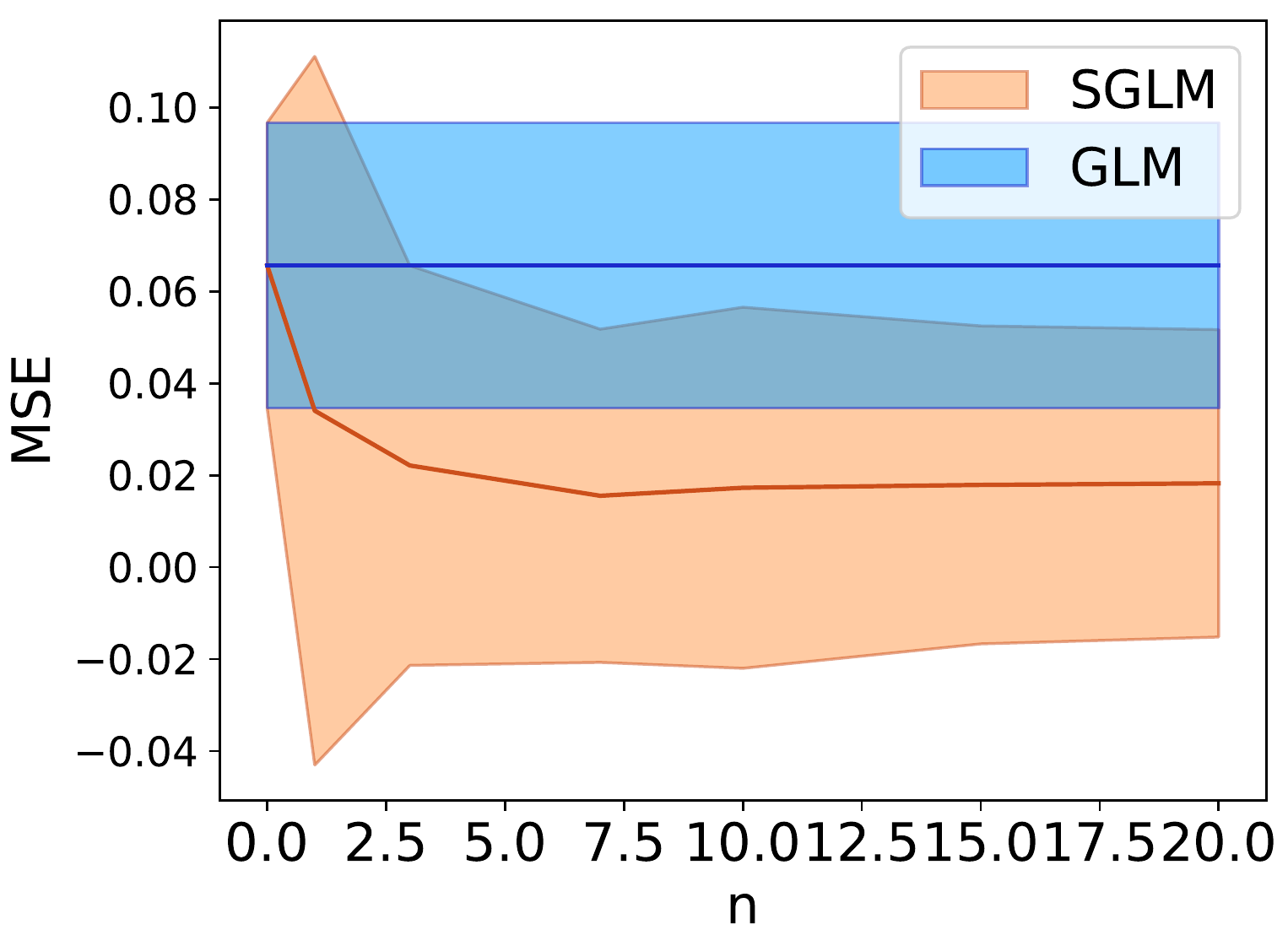}}
\subfigure[]{
  \includegraphics[width=0.4\textwidth]{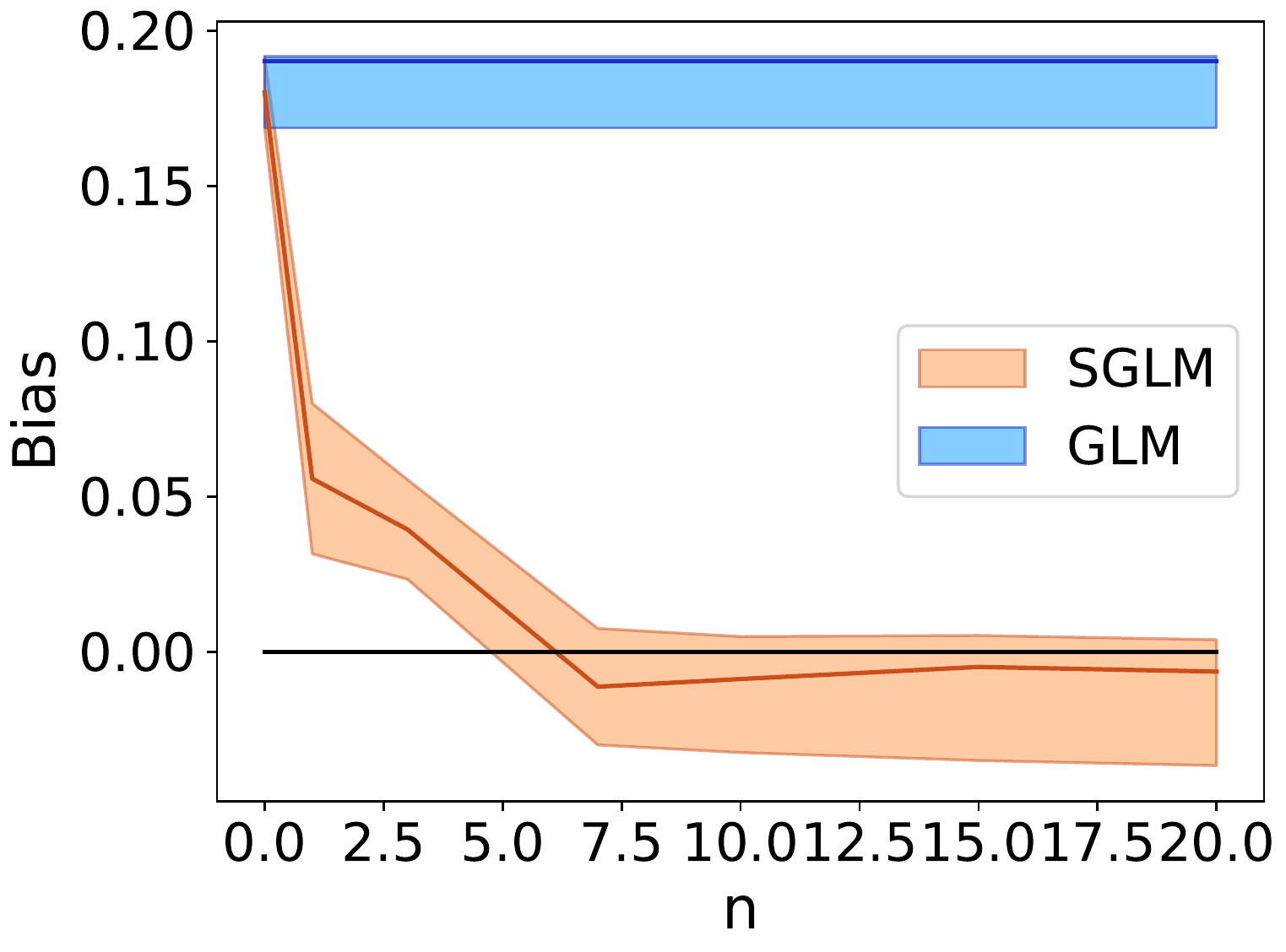}
  }
\subfigure[]{
  \includegraphics[width=0.4\textwidth]{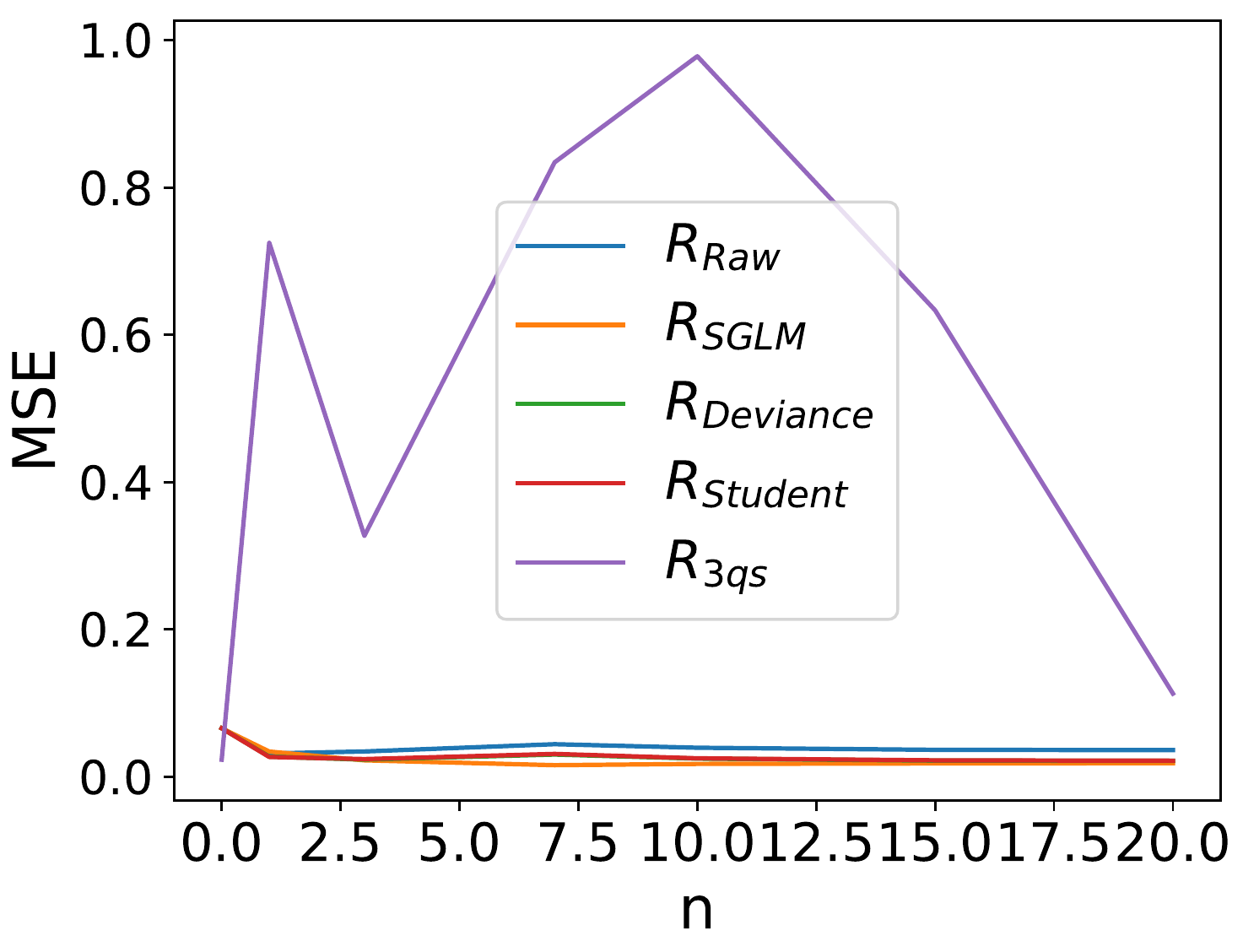}
}
\subfigure[]{
  \includegraphics[width=0.4\textwidth]{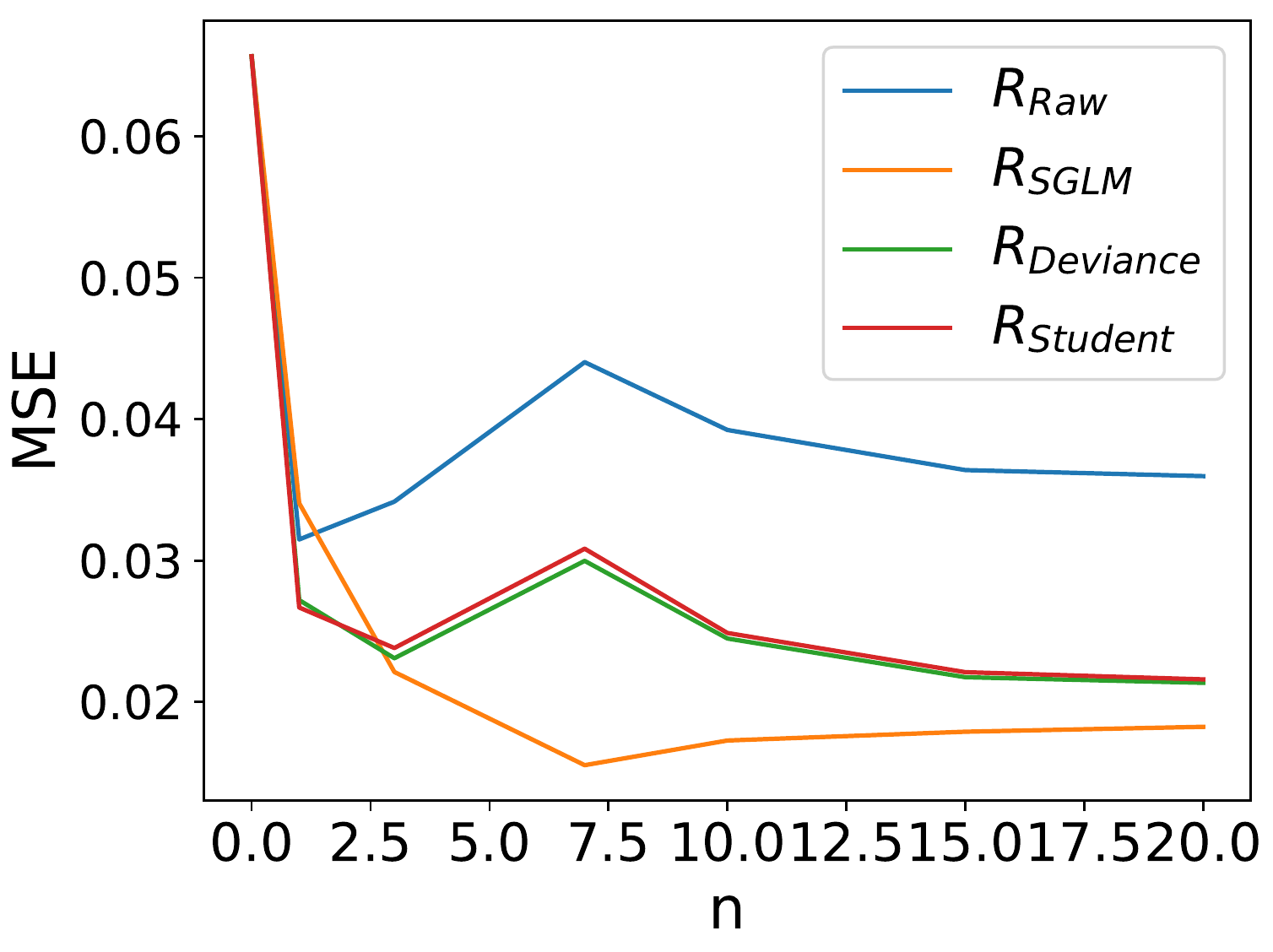}
}
  

\caption{Results on synthetic data. Mean Squared error (a), bias (b),  and residual comparison (c,d) vs dimension of $|Y_2|$ on Poisson regressions \label{fig:bias}}
\end{figure*}

Figure \ref{fig:bias} shows the estimation error for Poisson regression as a function of the dimension $n$ of $Y_{-1}$, i.e., the number of auxiliary variables available for denoising. In Figure \ref{fig:bias}(a) we plot the mean square error against the true value of $Z_1$ aggregated across the runs. Clearly increasing $n$ reduces error. This is expected, as the noise $N$ can be better estimated using more auxiliary variables. This in turn leads to lower error in estimates.
Due to the effect of $N$, the standard GLM estimates are biased. The simulation results bear this out, where we get more than \emph{10\%} bias in Poisson regression estimates. On the other hand, by being able to correct for the noise variable on a observation basis, our approach gives bias less than \emph{3\%}. We plot in Figure~\ref{fig:bias}(b) the bias of these estimates for Poisson. Once again as expected, increasing $n$ reduces the bias. 

We also experiment with other possible versions of residuals $\Rs$ \footnote{details in the Appendix} including residual deviance and student residuals. In Figure~\ref{fig:bias}(c) we have plotted the results of these methods against the method of \citet{3qs} (by fitting linear models on transformed observations). As is evident from the figure, data transformation, while a common practice, leads to substantially larger errors. Figure~\ref{fig:bias}(d) presents the effect of changing the residual definition. Under our interpretation of sibling regression (Claim \ref{clm:claim1}), any version of residual would be acceptable. This intuition is borne out it in these results as all the residuals perform reasonably. However from the figure, our proposed residual definition is the most effective.

\subsection{Discover Life Moth Observations}
Our next set of experiments use moth count surveys conducted under the Discover Life Project \footnote{https://www.discoverlife.org/moth}. The survey protocol uses screens illuminated by artificial lights to attract moths and then records information of each specimen. This data has been collected over multiple years at different locations on a regular basis.

\begin{figure*}[thb]
\centering
\subfigure[]{\label{fig:hfucosa}
\includegraphics[width=\columnwidth]{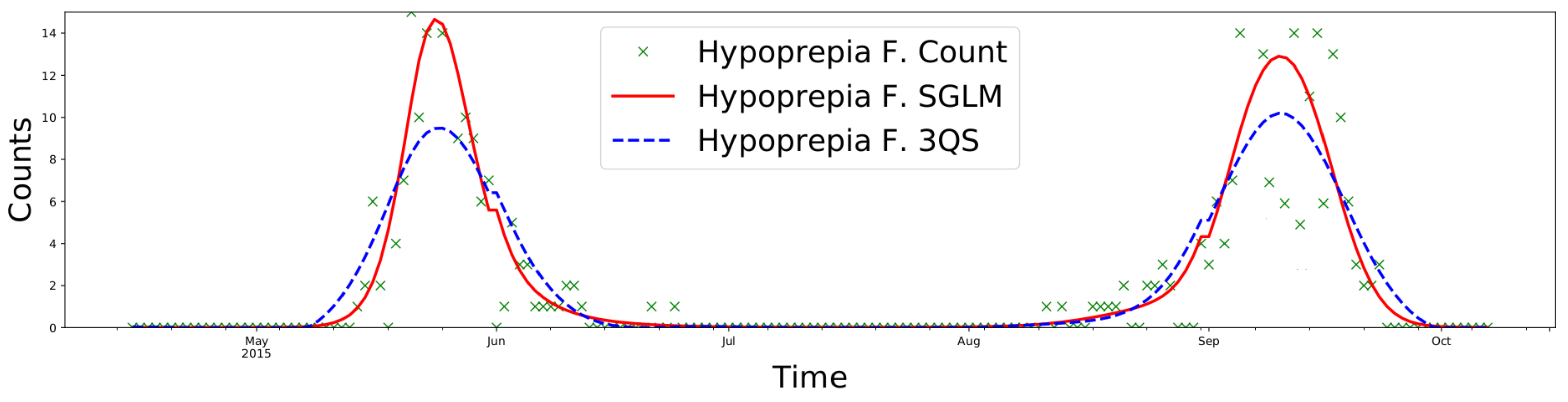}
}
\\
\subfigure{\label{fig:melegans}
\includegraphics[width=\columnwidth]{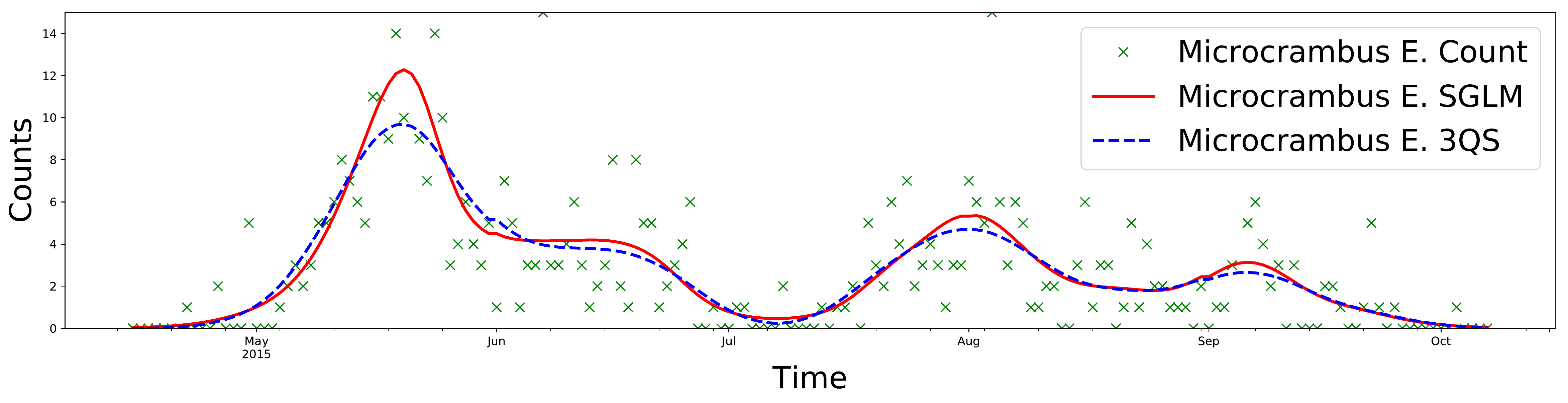}
}
\captionof{figure}{ \label{fig:glm_comp} Seasonal patterns of a) \emph{Hypoprepia fucosa} and b) \emph{Microcrambus elegans} as estimated by 3QS regression and our method alongside the observed counts. Note the higher peaks and better overall fit of our method.}
\end{figure*}

A common systematic confounder in such studies is moonlight. Moth counts are often low on full moon nights as the light of the moon reduces the number of moths attracted to the observation screens. In their paper, \citet{3qs} present the `three-quarter sibling' (3QS) estimator and use it to denoise moth population counts. However, to apply the model they used least-squares regression on transformed count variables. Such transformations can potentially induce significant errors in estimation. The more appropriate technique would be to build models via a Poisson generalized additive model (GAM). In this experiment, we use our technique to directly learn the underlying Poisson model.

\textbf{Description}
We follow the methodology and data used by \citet{3qs}. We choose moth counts from the Blue Heron Drive site for 2013 through 2018. We then follow the same cross-validation like procedure, holding each year out as a test fold, while using all other years for training. However, instead of transforming the variables, we directly estimated a Poisson GAM model with the pyGAM \citep{pyGAM} package. Next, we compute $Z_i$ with our SGLM-algorithm. This procedure is repeated for all folds and all species. We compare the MSE obtained by our estimates against the 3QS estimates. Note here that due to the absence of ground truth, the prediction error is being used as a proxy to assess the quality of the model.
The hypothesis is that correcting for systematic errors such as the ones induced by the moon will help to generalize better across years.

\textbf{Results}
First we compare the residuals as used in \citet{3qs} (by fitting linear models on transformed observations) against the residuals as obtained by our method, in terms of correlation with lunar brightness. A higher (magnitude) correlation indicates that the residuals are a better proxy for this unobserved confounder.
The results are in Table~\ref{tab:exp3}. For comparison, we also provide correlations  obtained by simply using the difference between the model prediction and observed values. Clearly, our method is most effective at capturing the effect of lunar brightness on the counts.

\noindent\begin{minipage}{\linewidth}

    \centering
          \begin{minipage}{0.45\linewidth}
    \includegraphics[width=0.9\linewidth]{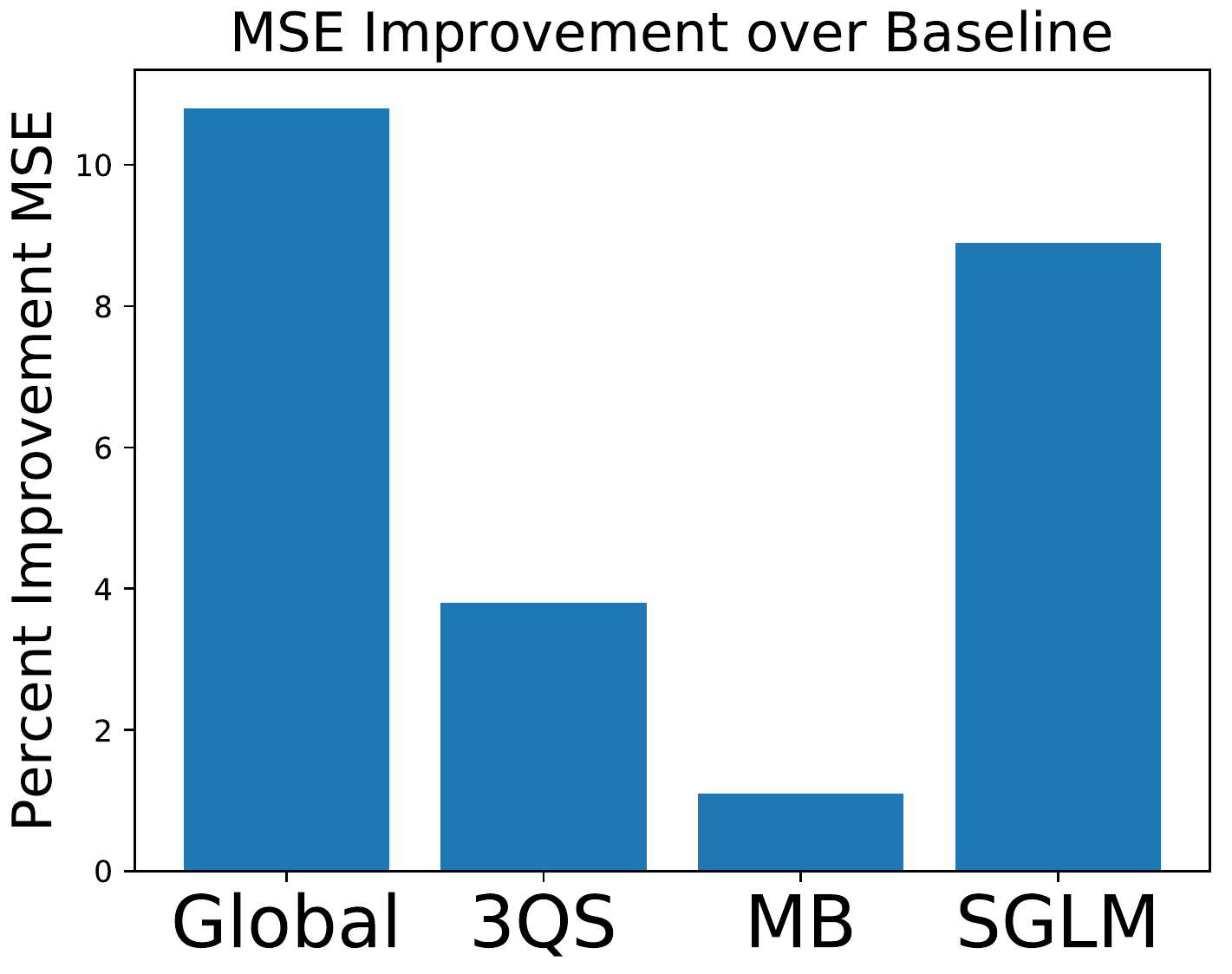}
  \captionof{figure}{Average percent improvement in predictive MSE relative to a GAM fitted to the raw counts \label{fig:mse_bar_chart}}
      \end{minipage}
      \begin{minipage}{0.45\linewidth}
  \small
\begin{tabular}{|c|crc|}
  \hline
  & \multicolumn{3}{|l|}{Moonlight Correlation} \\
Species & $R_{\text{SGLM}}$ & $R_{\text{3QS}}$ & $R_{raw}$ \\ 
\hline
\small Melanolophia C. & \textbf{ -0.55} & -0.21  & -0.42 \\
\small Hypagyrtis E. & \textbf{-0.66} &  -0.42 & -0.62 \\
\small Hypoprepia F. & \textbf{-0.65} &  -0.36 & -0.59 \\
\hline
\end{tabular}
\captionof{table}{\label{tab:exp3} Correlation with lunar brightness of different residuals. $\hat{R}_{\text{SGLM}}$ is our residual, $\hat{R}_{\text{3QS}}$ are residuals from 3QS estimator and $R_{Raw}$ is the raw difference between prediction and observation.}
      \end{minipage}
\end{minipage}

\textbf{}\\
Next, we compare the decrease in prediction error obtained by our method against the methods tested by \citet{3qs}. These results are presented in Figure~\ref{fig:mse_bar_chart}. The mean squared error (MSE) is computed only on
data from the test-year with moon brightness zero.
``Global'' is an oracle model shown for comparison. It is fit on   multiple years to smooth out both sources of year-to-year variability (intrinsic and moon phase). ``MB'' is a model that includes moon brightness as a feature to model detection variability.
From the figure, we can see that our method not only improves substantially over the 3QS estimator (9\% vs 4\%) but is comparable with the global model fit on multiple years (9\% vs 10\%).\footnote{The global model is included as a rough guide for the best possible generalization performance, even though it does not solve the task of denoising data within each year.} This is partly because the transformed linear model is unsuited for these variables. Our technique on the other hand can directly handle a broader class of conditional models and more diverse data types.

Finally, to present the difference in behavior of the two approaches, we plot the estimated moth population curves in Figure \ref{fig:glm_comp}. These curves are plotted for two different species with the dotted line representing the 3QS regression model, while the bold line represents our method. The actual observed counts are also plotted as points. One can clearly see the impact of the transformation, which produces flatter curves. For example, the height of the peaks for \textit{Hypoprepia fucosa} end up significantly lower than the observed counts. On the other hand, our method predicts higher and narrower peaks, which better match the observed values.

\section{Ethical Impact} 

\paragraph{Applications}
Our method is more focused towards ecological survey data applications.
Such surveys provide information useful for setting conservation policies. However there are other potential domains of application. Measurement noise is ubiquituous in experimental studies, and applied scientists often used different schemes to protect against confounding \cite{Genb_ck_2019} and measurement effects \cite{Zhang18}. As such our method may be useful applied to domains such as drug reporting~\cite{adams2019learning} and epidemiology~\cite{robin02}.

\paragraph{Implications} Our method provides an approach to handle the presence of unobserved confounding noise. The unobserved confounder however need not be a nuisance variable. Empirical datasets often exhibit different biases due to non-nuisance confounders \cite{davidson2019racial}, which can lead to unfairness with respect race, gender and other protected attributes \citep{Olteanu16, chouldechova18a}. In some cases a protected attribute itself might be a confounder, in which case our approach can have implications for developing fairer models. Since our method partially recovers the unobserved confounder (noise), it can lead to identification or disclosure of protected attributes even when such data has been hidden or unavailable. This could lead to issues regarding privacy and security. The proposed method does not handle such issues, and adequate measures may be  warranted for deploying this method..

\section{Conclusion}
Our paper has two primary contributions:
a) reinterpreting sibling regression in residual form which enabled the generalization to GLMs and b) presenting a residual definition which corresponds to the case of noise in natural parameters.
Based on these we designed a practical approach and demonstrated its potential on an environmental application.

A future line of work would be to develop goodness-of-fit tests for these models. A second question could be to generalize this chain of reasoning to complex non-linear dependencies. Finally since this method partially recovers the unobserved confounder (noise), it can potentially lead to identification of protected attributes even when such data has been hidden. As such another venue of future research is in the direction of how sibling regression can affect fairness and security of models. 



\end{document}